\documentclass[preprint,12pt,authoryear]{elsarticle}
\usepackage{amsmath}
\usepackage{amssymb}
\usepackage{amsthm}
\usepackage{mathtools}
\usepackage{multirow}
\usepackage{wrapfig}
\usepackage{subcaption}
\usepackage{array,multirow}
\usepackage{ulem}
\usepackage{lineno}
\usepackage{hyperref}

\newcommand{\azul}{\color{blue}}

\usepackage{tikz}
\usetikzlibrary[topaths]
\usetikzlibrary{arrows}
\usetikzlibrary{shapes.misc, positioning, arrows, decorations.text, shapes, mindmap,trees}

\date{\azul April 24$^{th}$, 2024}

\journal{Ecological Informatics}

\begin{document}


\newpageafter{author}

\begin{frontmatter}

\title{Bayesian feedback in the framework of ecological sciences}

\author[labelaff1]{Mario Figueira}
\ead{Mario.Figueira@uv.es}
\author[labelaff2]{Xavier Barber}
\author[labelaff1]{David Conesa}
\author[labelaff1]{Antonio López-Quílez}
\author[labelaff3]{Joaquín Martínez-Minaya}
\author[labelaff4]{Iosu Paradinas}
\author[labelaff5]{Maria Grazia Pennino}

\affiliation[labelaff1]{organization={Departamento de Estadística e Investigación Operativa. Universitat de València},
            country={Spain.}}

\affiliation[labelaff2]{organization={Centro de Investigación Operativa.Universidad Miguel Hernández de Elche}, country={Spain.}}

\affiliation[labelaff3]{organization={Departamento de Estadística e Investigación Operativa Aplicadas y Calidad, Universitat Politècnica de València}, county={Spain}}

\affiliation[labelaff4]{organization={Azti, Txatxarramendi Ugartea z/g, 
48395 Sukarrieta}, country={Spain}}

\affiliation[labelaff5]{organization={Instituto Español de Oceanografía (IEO-CSIC)}, country={Spain}}

\begin{abstract}

\begin{enumerate}
    \item In ecology we may find scenarios where the same phenomenon (species occurrence, species abundance, etc.) is observed using two different types of samplers. For instance, species data can be collected from scientific sampling with a completely random sample pattern, but also from opportunistic sampling (e.g., whale or bird watching fishery commercial vessels), in which observers tend to look for a specific species in areas where they expect to find it.
    \item Species Distribution Models (SDMs) are a widely used tool for analyzing this kind of ecological data. Specifically, we have two models available for the above data: {\color{blue} a geostatistical model} (GM) for the data coming from a complete random sampler and a preferential model (PM) for data from opportunistic sampling. 
    \item {\color{blue} Integration of information coming from different sources can be handled via expert elicitation and integrated models. We focus here in a sequential Bayesian procedure to connect two models through the update of prior distributions}. Implementation of the Bayesian paradigm is done through the integrated nested Laplace approximation (INLA) methodology, a good option to make inference and prediction in spatial models with high performance and low computational costs. This sequential approach has been evaluated by simulating several scenarios and comparing the results of sharing information from one model to another using different criteria. {\color{blue} The procedure has also been exemplified with a real dataset.}
    \item Our main results imply that, in general, it is better to share information from the independent (completely random) to the preferential model than the alternative way. However, it depends on different factors such as the spatial range or the spatial arrangement of sampling locations. 
\end{enumerate}

\end{abstract}


\end{frontmatter}

\section{Introduction}

Species Distribution Models (SDMs) are widely used in ecological analysis for various purposes such as inferring ecological niches \citep{Habitat_Distribution_Models_Guisan}, assessing climate change impacts on habitats \citep{ProjectingDistributionFisheryDependent_Karp}, predicting invaders and invasive species \citep{FutureInvadersSPDs_Fournier, HabitatManagementArthropod_Landis}, suggesting protected areas \citep{ParadinasDesigning_2021}, and refining biodiversity inventories \citep{BioticInteractionRefiningSPDs_Staniczenko}.


{\azul In this framework, it is common to have available multiple sources of information for the same ecological phenomenon}, like in fishery ecology where data can come from scientific surveys and/or commercial vessels \citep{Braun2023}. This raises the question of how to integrate information from different sources, {\azul which can be achieved in two different ways: integrated models and expert elicitation. }

{\azul
The integrated species distribution models (iSDM) represent a novel approach that involves combining various information sources to construct a unified model. In this approach, it is possible to jointly address diverse data, which may come from citizen science, scientific surveys, commercial fishing surveys, etc. This implies not only integrating different data but also integrating various sampling structures to reduce biases inherent in the sampling designs, and so, leveraging joint information to enhance inferential capacity \citep{Koshkina_IntegratedModels_2017, Fletcher_CominingDataSDM_2019}. In particular, \cite{Suhaimi_IntegratedModels_2021} propose various approaches for constructing a joint model that incorporates both presence-only and presence-absence data; \cite{rufener2021bridging}, \cite{Alglave_IntegratedModels_2022} and \cite{Paradinas_IntegratedModels_2023} also present ways for integrating independent-data and opportunistic or dependent-data into a single model analysis overcoming the bias and the scarcity of information derived from the particular isolated surveys or sample sets; and \cite{Jung_IntegratedModels_2023} describes \texttt{ibis.iSDM}, a modelling framework for iSMDs, that allows the integration of diverse data sources into a single model, supporting parameter transformations, tuning, and spatial-temporal projections.
}

{\azul In expert elicitation, experts provide their opinions and expertise to incorporate additional information in the analysis, mainly (but not exclusively) by defining prior distributions within the Bayesian framework. This involves multiple stages, including defining the format and model of the elicitation process and recruiting suitable experts \citep{UncertainJudgements_Haga, Elicitation_Dias}. Expert elicitation has demonstrated particular effectiveness in the realm of SDMs \citep{RedefiningExpertise_Burgman,vanhatalo2014catch, nevalainen2018estimating,lamere2020making, SPDElicitation_Kaurila}. In particular, \cite{crawford2020expert} employed expert opinion to shape habitat suitability models for conservation planning in the US; \cite{difebbraro2018expert} evaluated the feasibility of monitoring habitat quality for bird communities in central Italy using a blend of survey data and expert-driven models; and \cite{pearman2020predicting} leveraged expert elicitation to characterise the distribution of various species in New England, employing a web questionnaire to extract species presence probabilities and insights into covariate impacts on species occurrence.}


{\azul A particular case of expert elicitation is that of incorporating information from previous experiments/surveys via prior distributions. In other words, prior information can be extracted from estimates of similar effects from a different model to the one into which we wish to integrate this knowledge. This does not necessarily require the same data sources, but rather that the underlying process generating the phenomenon of interest should have a similar mathematical structure. The procedure, known as Bayesian feedback or prior elicitation, involves using the posterior distributions of the auxiliary model to inform the prior distributions of the parameters and hyperparameters in the model of interest.}


{\color{blue}
Our study proposes a protocol to perform a Bayesian feedback approach considering two types of sampling (i.e., an independent and a preferential), each of which is analysed with a specific model. In particular, independent sampling data focus on geostatistical processes \citep{Geostatistical_Diggle}, that can incorporate biotic \citep{BioticSPDs_Barber}, spatial and spatio-temporal effects \citep{BayesianSpatioTemporalNurseries_Paradinas, SPDsSpatioTemporal_MartinezMinaya}. Preferential sampling models (PM) are analysed using marked point pattern models which consider that the sampled quantities of the ecological phenomenon of interest (i.e., species occurrence or abundance) are influenced by the sampling process \citep{GeostatisticalPreferentialSampling_Diggle, FisheryDependentIndependent_Pennino, AccountingPreferentialSampling_Pennino}. Both models can use Bayesian hierarchical modelling, and prior information can be integrated through informative prior distributions. The prior elicitation trough the Bayesian feedback between these two models can be performed in both directions, from the IM to the PM or the other way around.}


{\color{blue}
Here, we present two methodologies for feedback between spatial Bayesian hierarchical models fitted with the Integrated Nested Laplace Approximation (INLA, \citealt{Rue2009, ReviewINLA_2017}) approach. The first methodology involves full Bayesian updating of the  marginal distributions, where the prior distributions of one model are replaced by the posterior distributions of the other. The second methodology updates the characteristic moments (mean, variance, mode, quantiles, etc.) of the prior distributions on one model based on the moments of the posterior distributions of the other model. 


In order to validate the two procedures we also present here a set of simulated scenarios through which we compare the different direction of the two feedback procedures. Evaluation and assessment of the behaviour of both procedures is done by means of the analysis of residuals from predictive maps, encompassing metrics like root mean square error, bias, histograms of residuals, and residual plots against predicted values. These simulated environments also allow us to identify possible biases in parameter and hyperparameter estimation. We finally present an application of the proposed method in the context of a real fishery scenario. In particular, we study the distribution of the European hake ({\it Merluccius merluccius}) in the southern French coast of the Bay of Biscay. In the analysis, we combine information gathered from fishery independent samples collected through the french EVHOE fishery trawl survey (FI samples), and fishery dependent samples collected through onboard sampling of basque pair trawlers.}


\section{Species distribution models}

SDMs are statistical models, in which biological data related to organism populations are linked to some explanatory variables. The response variable of the model, biological population data, can be either presence-only, presence/absence, proportional data, discrete abundance or continuous biomass data based on random or stratified field sampling, or observations obtained opportunistically \citep{Habitat_Distribution_Models_Guisan, PredictingSpeciesDistribution_Guisan, SPDs_EcologicalExplanation_Elith}. The kind of data related to explanatory variables can be either biotic (i.e. depredatory species distribution, tree cover density, ...) or abiotic (environmental data as soil data, temperature, salinity, ...), and these are chosen to reflect some of the three main types of effects on the species population data: (i) \textit{limiting factors} (or \textit{regulators}), (ii) \textit{disturbances} and (iii)  \textit{resources} \citep{PredictingSpeciesDistribution_Guisan}.

In our case, we will {\color{blue} work with two models, geostatistical (GM) and preferential (PM),} following the notation proposed in \cite{SPDsSpatioTemporal_MartinezMinaya}. {\color{blue} The {\color{blue} geostatistical model} comprises a predictor with three elements}: (i) an intercept, (ii) a linear effect for a spatial covariate, and (iii) a structured random spatial effect. On the other hand, the preferential model will include the same three elements but structured in two submodels: a geostatistical submodel, like the one used for the independent one, and a point process submodel for the sample structure {\color{blue} with an additional spatial effect that modifies the preferential sampling process}.

\subsection{{\color{blue} Geostatistical model}}

In general, a {\color{blue} geostatistical model} assumes that data are generated from a continuous spatial process and are constituted by the measurements of the phenomenon under study.

For the sake of simplicity, our {\color{blue} geostatistical model} will be formed by an intercept $\beta_0$, a linear effect $\beta_1$ for a spatial covariate {\color{blue} $X_i$ and a structured spatial random effect $\textbf{u}$}. {\color{blue} The response variable $Y_i$ will follow a Gamma distribution with mean $\mu_i$ and variance $\phi$}, to resemble a biomass distribution. Therefore, putting all these pieces together we have the structure as follows:
{\color{blue}
\begin{equation}
    \begin{array}{c}
         Y_i \sim \text{Gamma}(\mu_i,\phi),\\
         \log(\mu_i) = \beta_0 + \beta_1 \cdot X_i + u_i,\\
         \textbf{u}\sim N(\mathbf{0},\mathbf{Q}(\rho, \sigma)),\\
         \beta_0\sim N(0,\tau_0), \ \beta_1\sim N(0,\tau_1),\\
         \rho \sim f_{\rho}(\rho\mid\boldsymbol\alpha_\rho), \ \sigma\sim f_{\sigma}(\sigma\mid\boldsymbol\alpha_\sigma),
    \end{array}
    \label{eq:SimpleModelGeostatistic}
\end{equation}
}where $\tau_0$ and $\tau_1$ are the precision for prior distributions of the intercept and the linear effect of the covariate respectively, {\color{blue} and} $f_{\rho}$ and $f_{\sigma}$ are the prior distributions for the spatial range $\rho$ and the marginal standard deviation $\sigma$ of the spatial effect, {\color{blue} being} the characteristic parameters of these prior distributions, $\boldsymbol\alpha_\rho$ and $\boldsymbol\alpha_\sigma$, {\color{blue} known. We will devote a specific section to priors, establishing the different options that can be used for spatial effects}. {\color{blue} From now on, we will denote $\boldsymbol\theta$ as the complete set of fixed and random effects comprising the latent field, e.g. $\boldsymbol\theta = \{\beta_0, \beta_1, \mathbf{u}\}$, while $\boldsymbol\psi$ will denote the complete set of the hyperparameters related to the latent field and the likelihood, e.g. $\boldsymbol\psi = \{\tau_0, \tau_1, \rho, \sigma\}$.}

The simplicity of this structure will allow us to establish more clearly the variations in the fixed parameter estimates and to evaluate these differences according to the feedback mechanism.

\subsection{Preferential sampling}

{\color{blue} The second model that we will use here is a preferential model. Here, as for the {\color{blue} geostatistical model}}, we will have a series of locations $\mathbf{s}=\{s_1,..,s_n\}$, but while the {\color{blue} geostatistical model} assumes that these are random and do not share information with the marks of the sampling points, in the case of the preferential model the locations would be generated by a non-homogeneous Poisson process with intensity $\lambda$; i.e. log-Gaussian Cox process (LGCP) relative to the geostatistical marking process, which is why this kind of data is also called a \textit{marked point pattern} \citep{PointPatterns_Diggle}.

The preferential model can then be considered as a two-stage model that share information through some common components \citep{Krainski2018}. {\color{blue} In particular, if $Y_i$ represents the response variable of the quantity of interest (usually an abundance) and supposing it is Gamma distributed with mean $\mu_i$ and variance $\phi_i$, then the structure of the preferential model} can be expressed as follows:
{\color{blue}
\begin{equation}
\begin{array}{c}
Y_i|s_i \sim \text{Gamma}(\mu_i, \phi),\\
s_i\sim \text{LGCP}(\lambda_i),\\
\log(\mu_i) = \beta_0 + \beta_1 \cdot X_i + u_i,\\
\log(\lambda_i) = {\color{blue}\gamma}\cdot(\beta_0 + \beta_1 \cdot X_i + u_i) + {\color{blue}u^*_i}.
\end{array}
\label{eq:pref_model_structure}
\end{equation}
}

In this equation most of the parameters are already known from the {\color{blue} geostatistical model}, with the exception of the $u^*_i$ and the ${\color{blue} \gamma}$ parameter. {\color{blue} The $u^*_i$ represents a specific spatial effect associated with the generating location process. It enables the consideration of other spatially structured elements that may influence the point process but are not directly linked to the response variable of the geostatistical process. Meanwhile, $\gamma$ is a parameter introduced by INLA during the process of incorporating the effect from the geostatistical likelihood. This parameter facilitates establishing a linear scaling in the sharing components} \citep{BayesianInferenceINLA_GomezRubio}. {\color{blue} In our case, this scale refers to the transformation of the intensity of the coefficients of the covariates and the spatial effect with respect to the geostatistical process, while maintaining the same spatial range of the spatial effect.}

\section{Inference}

{\color{blue}
The analysis of the raised issue was carried out through Bayesian hierarchical models. Moreover, using the INLA methodology implemented in the \texttt{R-INLA} software which has become in a well-established tool for Bayesian inference in many research fields \citep{Blangiardo2015, Lindgren2015}, including ecology \citep{BayesianSpatioTemporalINLAFishery_Cosandey, AccountingPreferentialSampling_Pennino, SpatioTemporalModelStructure_Paradinas}, epidemiology \citep{Blangiardo2013, GeospatialHealthData_Moraga} and econometrics \citep{SpatialEconometricsINLA_GomezRubio} due to its versatility and high performance. This will allow us to perform protocols for sharing information between models. In particular, to evaluate the spatial structure the SPDE approach is applied jointly with \textit{finite element methods} (FEM) allowing a fast and low cost computational resolution of spatial latent effects.}

\subsection{Normal transformed priors and PC-priors}

In INLA there are two {\color{blue} main} ways to write the prior distributions of the spatial effect. One using {\azul exponential transformations of normal distributions} (EN-priors), and the other using penalized complexity distributions (PC-priors).

The first type of prior distributions involves defining an initial value for the range and variance and the precision of a null-mean normal distribution that is exponentially transformed \citep{Lindgren2015}:
{\color{blue}
\begin{equation}
\begin{array}{c} 
     \rho=\rho_0\cdot \exp(\theta_1), \quad \theta_1 \sim N(0, 1), \\
     \sigma=\sigma_0 \cdot \exp(\theta_2), \quad \theta_2 \sim N(0, 1). \\
\end{array}
\label{eq:ENpriorSpatialEffect}
\end{equation}
}

{\color{blue} Therefore, as the mean is fixed to zero (which implies that the mode of $p(\rho)$ and $p(\sigma)$ are $\rho_0$ and $\sigma_0$ respectively) and the normal distribution is transformed by a exponential,} we have a positive definite distribution for the hyperparameters of the spatial effect. Moreover, this allows us to define an uninformative prior by setting a low precision while preserving the positive definite condition. But the posterior distributions of these hyperparameters would have a notable interpretative deficiency, since they would be provided with respect to the characteristic parameters of the Normals. {\color{blue} As a result, direct interpretation of the hyperparameters ($\rho$ and $\sigma$) of the spatial effect is not feasible unless we perform the exponential transformation illustrated in equation (\ref{eq:ENpriorSpatialEffect}).}

The second type of prior distributions was designed precisely to avoid this lack of interpretability. Hence, the penalized complexity prior distributions proposed by \cite{Simpson2017}, and extended in \cite{HyperpriorPreferentialModel} and \cite{Fuglstad2019}, are defined by two elements: (i) a fixed $\rho_0$ or $\sigma_0$ value, and (ii) a probability value  $p_0$ which indicates the probability above or below that ($\rho_0$, $\sigma_0$) initial value. Thus, the prior distribution for $\rho$ and $\sigma$ are define as follows
\begin{equation}
\begin{array}{c}
    \rho \sim \text{PC-prior}(\rho_0, p_{0}), \\
    \sigma \sim \text{PC-prior}(\sigma_0, p_{0}),
\end{array}
\label{eq:PCpriors1}
\end{equation}
where this $\text{PC-prior}$ differs in the definition of tail probability depending on whether it refers to the spatial range ($\rho$) or to the marginal standard deviation ($\sigma$). {\color{blue} In particular,}
\begin{equation}
\begin{array}{c}
    \text{PC-prior}(\rho_0, p_{\rho0})\equiv P(\rho<\rho_0)=p_0, \\
    \text{PC-prior}(\sigma_0, p_{\sigma0}) \equiv P(\sigma>\sigma_0)=p_0.
\end{array}
\label{eq:PCpriors2}
\end{equation}
{\color{blue} For instance, defining $p_0=0.5$ means that the probability of $\rho$ being less than $\rho_0$ and $\sigma$ being greater than $\sigma_0$ are equal to $0.5$, setting the median of the prior distribution.}


\section{Bayesian Feedback}

{\color{blue} In what follows we introduce our proposal to perform a Bayesian feedback procedure within the context of Species Distribution Models. The basic scheme starts with the fitting of model $M_1$ (either a geostatistical or preferential model) with data $\mathbf{y}_1$, obtaining posterior distributions of parameters and hyperparameters $\pi(\boldsymbol\theta,\boldsymbol\psi|M, \mathbf{y}_1)$, and then using them to feed back the inferential process of fitting another model $M_2$ (the corresponding contrary model considered in $M_1$) for a new data set $\mathbf{y}_2$. For instance, when analyzing the distribution of the European hake {(\it Merluccius merluccius}) in the southern French coast of the Bay of Biscay, as later described, we could first analyze the information gathered from fishery independent samples collected through a trawl survey, and then using the posterior distribution of the parameters of the geostatistical model to feed back the fishery dependent samples collected through onboard sampling of Basque pair trawlers. Or viceversa.}
 
{\color{blue} In the remaining of this Section, we first introduce two different updating protocols, and then the two possible feedback situations we can find when having two different models, that is, feed-backing a geostatistical model with information from a preferential model and viceversa.}

\subsection{\color{blue} Updating structures for feed-backing}

{\color{blue} Figure \ref{fig:SingularModelBayesianFeedback} schematically presents two proposals for performing Bayesian feedback:
\begin{enumerate}
    \item[\textbf{(i)}] \textbf{Full updating:} this protocol just implies the replacement of the prior distributions by the posterior distributions. It is the simplest feedback procedure conceptually and naturally in Bayesian statistics. 
    \item[\textbf{(ii)}] \textbf{Updating by moments:} this approach assumes that prior distributions can be analytically defined by updating the characteristic parameters of the kernel in the posterior distributions with their estimates from the prior. For example, if a distribution is characterised by its mean and variance, both parameters could be updated with the corresponding values from the posterior distribution. This method is suggested because the INLA structure requires latent field distributions to conform to a Gaussian field. In addition, the hyperparameter distributions are internally re-parameterised to resemble Gaussian distributions, which allows us to easily update the hyperparameters in the internal parameterisation.
\end{enumerate}
}

{
\linespread{1.}
\begin{figure}[h!]
    \centering
    \begin{tikzpicture}
    \node at (-1,0) [circle,draw] (1) {$\mathbf{y}$};
    \node at (-1,-3) [circle,draw] (2) {$\everymath={\displaystyle}
    \begin{array}{c}
    \boldsymbol\alpha_i \\[0.2cm]
    \boldsymbol\alpha_j\\
    \end{array}$}; 
    
    \node at  (3,-1.5) [rectangle,draw] (3) {$\everymath={\displaystyle}
    \begin{array}{c}
    \mathbf{Y} \sim  f(\mathbf{y}|\boldsymbol\eta,\boldsymbol\psi)\\[0.2cm]
    g\left[E(\mathbf{Y})\right]  =  \eta(\mathbf{x}|\boldsymbol\theta,\boldsymbol\psi)\\[0.2cm]
    \theta_i \sim \pi_i(\theta_i|\boldsymbol\alpha_i)\ \\[0.2cm]
    \psi_j \sim \pi_j(\psi_j|\boldsymbol\alpha_j)\ \\
    \end{array}$};
    
    \node at  (7.5,-1.5) [rectangle,draw] (4) {$\everymath={\displaystyle}
    \begin{array}{c}
    \pi_i(\theta_i|\mathbf{y},\boldsymbol\alpha_i,\boldsymbol\psi_i)\ \\[0.2cm]
    \pi_j(\psi_j|\mathbf{y},\boldsymbol\alpha_j)\ \\
    \end{array}$};
    
    \node at (4,-4.5) [circle,draw] (5) {$\everymath={\displaystyle}
    \begin{array}{c}
    \boldsymbol\alpha_i^*, \ \boldsymbol\alpha_j^*
    \end{array}$};
    
    \node at (3,1.25) [circle,draw] (6) {$\mathbf{y}^{new}$};
    
    \node at (10.5,-1.5) [circle,draw] (7) {$\mathbf{y}^{pred}$};
    
    \draw[-to] (1) to [out=-90,in=170,looseness=1.25] (3);
    \draw[-to] (2) to [out=90,in=190,looseness=1.25] (3);
    \draw[double,-to] (3) to (4);
    \draw[dotted,-to] (4) to [out=270,in=0,looseness=1.5]  (5);
    \draw[dotted,-to] (4) to [out=90,in=40,looseness=1.5] (3);
    \node[text width=4cm,fill=white] at (8.5,-.1) {\textbf{(i)} \textit{Full updating}};
    \draw[dotted,-to] (5) to [out=-180,in=-20,looseness=1.5] (2);
    \node[text width=8cm,fill=white] at (10.5,-3.3) {\textbf{(ii)} \textit{Updating by moments}};
    \draw[dotted,-to] (5) to [out=-180,in=-20,looseness=1.5] (2);
    \draw[dotted,-to] (6) to [out=-180,in=40,looseness=1.5] (1);
    \draw[double,-to] (4) to (7);
    \end{tikzpicture} 
    \caption{\color{blue} Illustration of the two proposals for incorporating information gathered from a previous fitting in a new analysis.}
    \label{fig:SingularModelBayesianFeedback}
\end{figure}
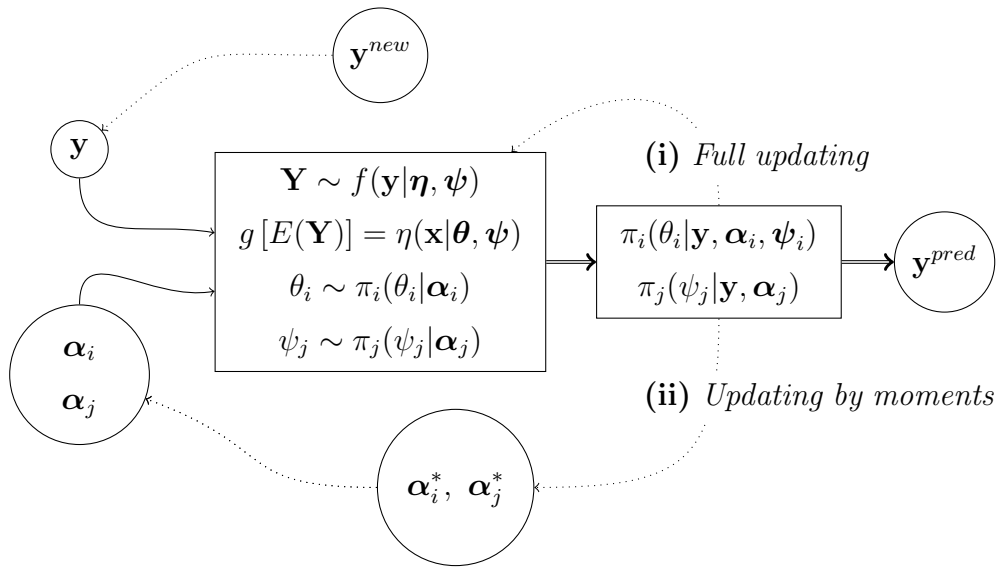
}

{\color{blue} However, it is worth noting that regardless of which of the two approaches we employ, it is not possible to simultaneously report the random effects of the latent field and the hyperparameters associated with the prior distributions of these random effects. Therefore, in this study we do not consider at any point incorporating information on the random effects of the latent field, only information on the fixed effects and the hyperparameters of the latent field. } 

{\color{blue} In any case, being our focus describing the two above mentioned possible feedback situations (feed-backing a geostatistical model with information from a preferential model and viceversa), Figure \ref{figure:EssentialComponents} jointly describes the two datasets $\mathbf{y_G}$ and $\mathbf{y_P}$, the two model structures $M_G$ and $M_P$, and the posterior distributions $\pi(\boldsymbol\theta,\boldsymbol\psi|M_G,\mathbf{y_G})$ and $\pi(\boldsymbol\theta,\boldsymbol\psi|M_P,\mathbf{y_P})$.
}

{
\linespread{1.}
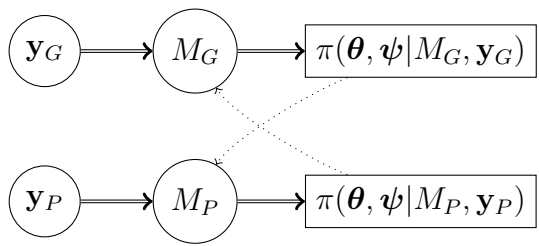
\begin{figure}[h!]\centering
    \begin{tikzpicture}
  
    \node at (-1,-4) [circle,draw] (3) {$\mathbf{y}_G$};
    \node at (1,-4) [circle,draw] (4) {$M_G$};
    \node at (4,-4) [rectangle,draw] (5) {$\pi(\boldsymbol\theta,\boldsymbol\psi|M_G,\mathbf{y}_G)$};

    \node at (-1,-6) [circle,draw] (7) {$\mathbf{y}_P$};
    \node at (1,-6) [circle,draw] (8) {$M_P$};
    \node at (4,-6) [rectangle,draw] (9) {$\pi(\boldsymbol\theta,\boldsymbol\psi|M_P,\mathbf{y}_P)$};

    \draw[double,-to] (3) to (4);
    \draw[double,-to] (4) to (5);
    \draw[dotted,-to] (5) to [out=200,in=60,looseness=0.5]  (8);
    \draw[double,-to] (7) to (8);
    \draw[double,-to] (8) to (9);
    \draw[dotted,-to] (9) to [out=-200,in=-60,looseness=0.5]  (4);
    \end{tikzpicture} 

    \caption{\color{blue} Essential components and scheme for Bayesian feedback between geostatistical and preferential models.}
    \label{figure:EssentialComponents}
\end{figure}
}

The main distinction between the two schemes is that, in this case, the feedback occurs between two distinct models. As a result, the feedback process is asymmetric, as the {\color{blue} geostatistical model} and the preferential model have different parameters and hyperparameters.

\subsection{{\color{blue} Geostatistical model} feedback}

In the first place, we will study the feedback of the {\color{blue} geostatistical models} by assuming given the fit of the preferential model. As we have indicated above with respect to the feedback schemes, the feedback will be performed by replacing the prior distributions by the posterior ones or by updating the characteristic parameters of the prior distributions according to the estimation of the same from the posteriors of the common parameters and hyperparameters. 

Therefore, considering that the parameters and hyperparameters of the {\color{blue} geostatistical model} and those of the preferential model, it is evident that the feedback relationship between the two is asymmetric, since starting from the PM all the parameters and hyperparameters of the IM can be updated, but not the other way around. In the Eq. \ref{eq:ParametersHyperparametersIMPM} we can compare the whole set of parameters and hyperparameters.

\begin{equation}
    \begin{array}{c}

    \text{GM}\equiv \left\lbrace\begin{array}{c}
    \boldsymbol\theta \\ \hline 
    \beta_0\sim N(0,\tau_{(\beta_0)}),\\
    \beta_1\sim N(0,\tau_{(\beta_1)}).\\ \hline \hline
    \boldsymbol\psi \\ \hline 
    \rho\sim \text{PC-prior}(\rho_0, p_\rho),\\
    \sigma\sim \text{PC-prior}(\sigma_0, p_{\sigma}),\\
    \log(\phi) \sim \log-\text{Gamma}(\mu_{(\phi)}, \phi_{(\phi)}).\\
    \end{array}\right\rbrace,
    \\ \\
    \text{PM}\equiv \left\lbrace\begin{array}{c}
    \boldsymbol \theta \\ \hline
    \beta_0\sim \text{N}(0,\tau_{(\beta_0)}), \\
    \beta_1\sim \text{N}(0,\tau_{(\beta_1)}), \\ \hline \hline 
    \boldsymbol \psi \\ \hline 
    \rho\sim \text{PC-prior}(\rho_0, p_\rho),\\
    \sigma\sim \text{PC-prior}(\sigma_0, p_{\sigma}),\\
    \alpha\sim \text{N}(0, \tau_{(\alpha)}),\\
    \log(\phi) \sim \log-\text{Gamma}(\mu_{(\phi)}, \phi_{(\phi)}).\\
    \end{array}\right\rbrace .
    \end{array}
\label{eq:ParametersHyperparametersIMPM}
\end{equation}

In short, the feedback process entails updating the set of fixed parameters $\{\mu_{(\beta_0)}, \tau_{(\beta_0)},\mu_{(\beta_1)}, \tau_{(\beta_1)}, \rho_0, p_{\rho},\sigma_0,p_{\sigma}, \mu_{(\phi)}, \phi_{(\phi)}\}$ with their estimated analogues from the posterior distributions of the latent effects and PM hyperparameters, common in GM. {\color{blue} Regarding the prior distributions of the hyperparameters of the spatial effect, we will employ PC-priors for the base models (the model without any feedback) as they allow easy definition of vague prior distributions. However, for the feedback, we will define normal distributions for the logarithmic transformation of these hyperparameters, as they are more flexible to define informative prior distributions:}
\begin{equation}
\begin{array}{c}
     \log(\rho)\sim \log(\rho_0) + \text{N}(0,\tau_\rho),\\
    \log(\sigma)\sim \log(\sigma_0) + \text{N}(0,\tau_\sigma).
\end{array}
\end{equation}
{\color{blue} However, when applying this feedback approach, one may encounter identification challenges. This is attributed to the notable discrepancy in gamma accuracy between the geostatistical and preferential models, particularly evident with a limited quantity of samples.}

\subsection{Preferential model feedback}

{\color{blue}
Now we present how to perform the feedback of the {\color{blue} preferential models} by assuming given the fit of the geostatistical model. As previously noted, this feedback mechanism differs from that of the geostatistical model. Specifically, this implies that it would not be feasible to directly incorporate all parameters associated with the likelihood of the point process or the hyperparameter $\gamma$ solely from the results of the geostatistical model. Consequently, to integrate the whole set of parameters and hyperparameters into the feedback loop, it would become necessary to conduct a point process fitting, yielding the latent effects and the hyperparameters of range and marginal standard deviation. For $\gamma$, an estimate could be established assuming normality and employing uncertainty propagation, typically through the Laplace approximation (refer to Appendix). However, since this approach doesn't seem to enhance adjustment outcomes significantly, we've opted to automate the process of integrating feedback from the preferential process, thus bypassing the need for an additional fitting procedure. In essence, the feedback process of the preferential model involves incorporating all parameters of the geostatistical model, except for the random effects of the latent field, as mentioned earlier.}

\section{\color{blue} A simulation study of the protocols}

{\color{blue} In this section, we present a set of simulated scenarios through which we compare the behaviour of the different direction of the two feedback procedures. In particular, we first provide a scheme of how to perform the simulation of the biomass/abundance, and then how to mimic the real sampling processes by sampling from the simulated scenarios. For a more detailed explanation of the simulation, please refer to the Appendix.}

\subsection{\color{blue} Spatial abundance/biomass simulation}

{\color{blue} In order to simulate a spatial abundance/biomass simulation with a latent spatial process, the initial step involves simulating a continuous covariate ($X_i$), multiplying it by its linear coefficient $\beta_1$, and then adding the latent spatial effect ($u_i$) in a study region defined as a square $10 \times 10$. These components are then combined with the global mean effect ($\beta_0$) to form the linear predictor. Subsequently, data are simulated following a Gamma distribution, aligning with biomass or abundance distribution scenarios, where the mean corresponds to the exponential of the sum of these effects. Additionally, the precision of the Gamma ($\phi$) is set sufficiently high to mitigate significant variability from the data distribution, preventing it from overshadowing the structure of the linear predictor. The mathematical structure of the model for simulation is then: 
\begin{equation}
\begin{array}{c}
    Y_i \sim \text{Gamma}(y_i \mid \mu_i, \phi),  \\
    \log(\mu_i)  = \beta_0 + \beta_1 \cdot X_i + u_i.
\end{array}
\label{eq:simulation_Geostatistic}
\end{equation}

In Table \ref{tab:simulation_Geostatistic}, we provide the values of the parameters and hyperparameters used to define the several scenarios. These scenarios were generated by considering all possible combinations of these values along with those related to the sampling schemes. Additionally, for each resulting scenario, we conducted 10 simulation replicates.}

\begin{table}[h!]
    \centering
    \begin{tabular}{|c|c|c|c|c|} \hline
        $\beta_0$ & $\beta_1$ & $\rho$ & $\sigma$ & $\phi$ \\ \hline
        $-1$ & $2$ & $(0.2, 0.5, 0.8)$ & $(0.5, 1)$ & $10$ \\ \hline
    \end{tabular}
    \caption{Values for the parameters and hyperparameters to simulate the different scenarios.}
    \label{tab:simulation_Geostatistic}
\end{table}

\subsection{\color{blue} Sampling from the spatial abundance/biomass field}

{\color{blue} Once we have the simulated scenarios, we have to mimic the usual sampling procedures (independent and preferential):}

\begin{enumerate}
\item[1.] \textbf{Independent sampling}. {\color{blue} Within the region of study, we conducted a uniform random generation for both the $X$ and $Y$ dimensions.}

\item[2.] \textbf{Preferential sampling}. {\color{blue} The simulation of preferential samples was conducted by defining the intensity of a LGCP along the study region. To replicate the structure of the preferential model outlined in Eq. \ref{eq:pref_model_structure}, we scaled the geostatistic linear predictor and introduced a spatial effect ($u^*$) specific to sample generation: $\lambda = \exp\left[{\color{blue}\gamma} \cdot (\beta_0 + \beta_1\cdot x + u) + u^*\right]$, where we set $\gamma$ to $0.5$ across the different scenarios. However, since we needed to control the number of generated samples —a random quantity in the LGCP— we optimized the total expected number of points $\Lambda$ through the following objective function, in which we integrate the new element $a$ to control the expected number of points within the study region ($\Omega$):

\begin{displaymath}
\def\arraystretch{2.2}
\begin{array}{lcr}
   \displaystyle \Lambda = \int_{s\in \Omega}\lambda \cdot ds & = & \displaystyle \int_{s\in\Omega}\exp\left[{\color{blue}\gamma}\cdot(\beta_0 + \beta_1\cdot x + u) + u^* + a\right]ds \\
     & \approx & \displaystyle\sum\exp\left[{\color{blue}\gamma}\cdot(\beta_0 + \beta_1\cdot x_i + u_i) + u^*_i + a\right] \Delta,
\end{array}
\end{displaymath}
where $\Delta$ is the area related to a minimum size for each data location generated along the study region $\Omega$. This last step is done given the need to discretize the space for computationally generate the samples from the LGCP.}
\end{enumerate}

\subsection{\color{blue} Set of scenarios}


{\color{blue} Finally, we have to decide how would be the different scenarios be in order to evaluate the performance of the feedback processes. In particular, we have used two types of designs for the samples: a balanced design (with equal quantities for independent and preferential sampling) and an unbalanced or asymmetric design (where one of the samplings in a case has a large amount of data and the alternative sampling has one of the various small sizes that have been stated). For each design, the different sample sizes chosen are presented in Table \ref{tab:values_sampling}.

In the balanced design, it must be considered that the primary concern is for the samples to be balanced, so these quantities are the expected values for the LGCP process and not the definitive values. That is, the quantity of samples for the preferential process defined as an LGCP is a random quantity whose expectation we can control, so once the quantity of samples following the previously explained procedure is simulated, we perform the simulation of the same quantity of samples for the independent sampling. In conclusion, regarding balanced samples, the final number of scenarios is the combination of the parameters that characterise the response variable along with the size of samples and the ten replicas for each of these combinations, assuming around $240$ scenarios.

In the unbalanced design, the primary concern is to compare scenarios in which there are two very different samples, one with little information and another with a lot of information. Therefore, the procedure followed has been to perform the combination of the parameters that configure the response variable with the different possible values for the small samples for one of the two types of sampling, fixing the larger sample for the other. The same procedure was followed for the alternative sampling design, considering that ten replicas have also been drawn for each of these combinations. This results in $360$ different scenarios.
}

\begin{table}[h!]
    \centering
    \begin{tabular}{|c|c|c|} \hline
        Symmetric   &  \multicolumn{2}{c|}{$50, 100, 250, 1000$}\\ \hline \hline 
                    & Small & Huge \\ \hline 
        Asymmetric  &  $50, 75, 150$ & $1000$ \\ \hline 
    \end{tabular}
    \caption{\color{blue} Set of scenarios induced by the different sample sizes selected both under the symmetric and asymmetric schemes.}
    \label{tab:values_sampling}
\end{table}

\section{Results}

{\color{blue} In this section, we present the results of the four fitting structures across the different scenarios for both geostatistical and sampling processes. In other words, we consider the proposed protocols as they were explained in the previous sections: using the updating by moments for all the parameters and hyperparameters with exception of the precision of the gamma distribution, which is update through the full updating protocol. In order to obtained these results we have used the PC priors in the base models for the spatial hyperparameters. Then we have updated these distributions in the feedback models using the normal distribution for the re-parameterisation of the spatial hyperparameters.}

{\color{blue} In what follows, we firstly illustrate the quality of the approximation in the updating by moments protocol and the change in the posterior distribution between the base model and its feedback counterpart. Additionally, we analyse the distributions of the updated parameters and hyperparameters. Then we show how three quantiles are distributed alongside the replicas for the parameters and hyperparameters. Finally, we present quantitative results from the analysis by means of out-of-sample validation, e.g. the mean global RMSE and bias for the four models (the two base models and their feedback counterparts).}

\subsection{Updating by moments vs reference posteriors}

\begin{figure}[h!]
    \centering
    \begin{subfigure}{0.475\linewidth}
        \includegraphics[width=\linewidth]{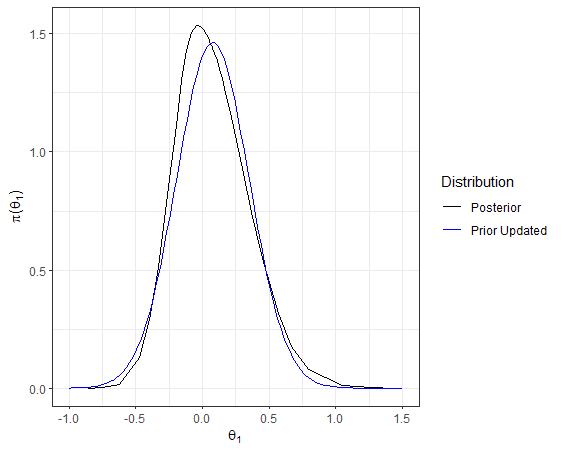}
        \caption{$\theta_1$ posterior and prior.}
    \end{subfigure}\hfill\vspace{5mm}%
    \begin{subfigure}{0.475\linewidth}
        \includegraphics[width=\linewidth]{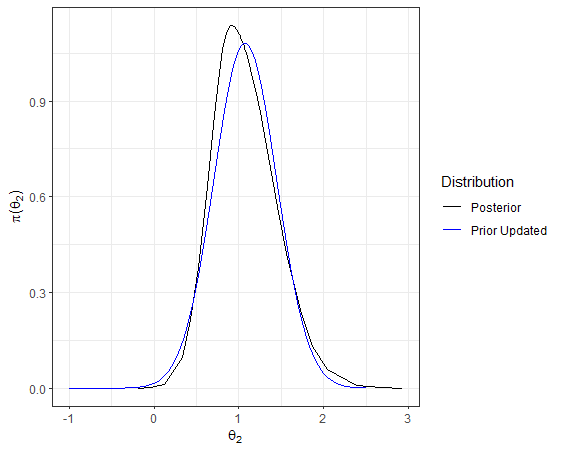}
        \caption{$\theta_2$ posterior and prior.}
    \end{subfigure}\hfill%

    \begin{subfigure}{0.475\linewidth}
        \includegraphics[width=\linewidth]{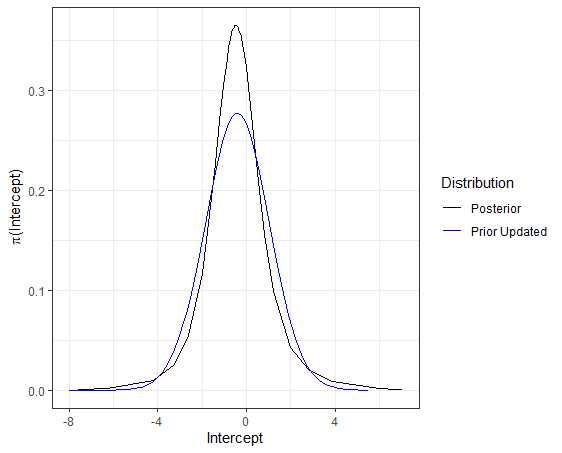}
        \caption{Intercept posterior and prior.}
    \end{subfigure}\hfill\vspace{5mm}%
    \begin{subfigure}{0.475\linewidth}
        \includegraphics[width=\linewidth]{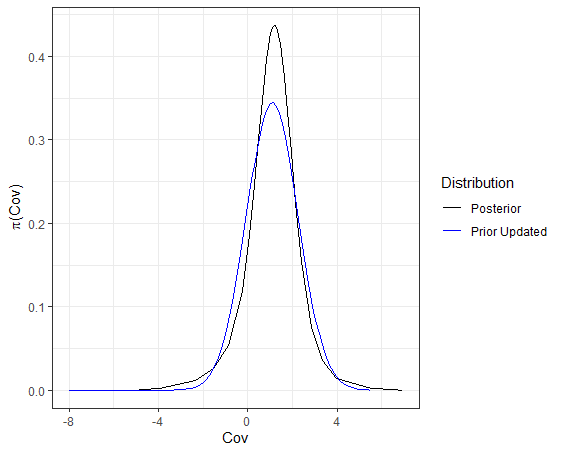}
        \caption{Covariate posterior and prior.}
    \end{subfigure}\hfill%
    \caption{\color{blue} Posterior and prior distributions. The prior distributions are defined by updating the moments estimated in the posterior.}
    \label{fig:PriorConstructedVSPosterior}
\end{figure}

{\color{blue} The first protocol (full update) clearly replicates information from the posterior distribution to the prior, essentially creating an exact copy. Conversely, the second protocol (update by moments) raise some concerns, as it updates characteristic parameters of a kernel distribution, which does not guarantee similarity between the distributions. However, the results indicate that distributions constructed by updating their moments closely resemble the reference posteriors from which the parameters are estimated. This similarity is illustrated in the graphs in Figure \ref{fig:PriorConstructedVSPosterior}, where we observe the close resemblance between the two distributions.}

\subsection{\azul Posteriors comparison between base and feed back models}

Feedback's influence on posterior distributions varies when assessing either fixed parameters or hyperparameters. {\color{blue} In general, the analysis of the posteriors reveals an improvement in parameter identification, resulting in reduced variances and enhanced accuracy compared to the values used in the simulation. This improvement is particularly notable when prior information is provided about the fixed parameters.} 

\begin{figure}[h!]
    \centering
    
    \begin{subfigure}{0.5\linewidth}
        \includegraphics[width=\linewidth]{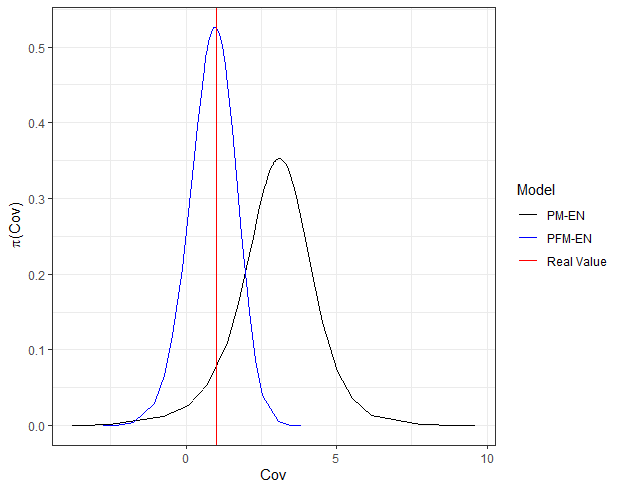}
        \caption{Covariate posteriors for PM with EN priors.}
    \end{subfigure}\hfill%
    \begin{subfigure}{0.5\linewidth}
        \includegraphics[width=\linewidth]{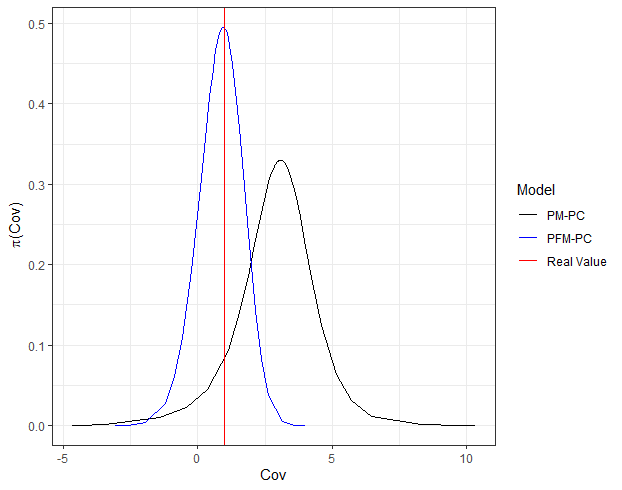}
        \caption{Covariate posteriors for PM with PC-priors.}
    \end{subfigure}\hfill%
    
    \begin{subfigure}{0.5\linewidth}
        \includegraphics[width=\linewidth]{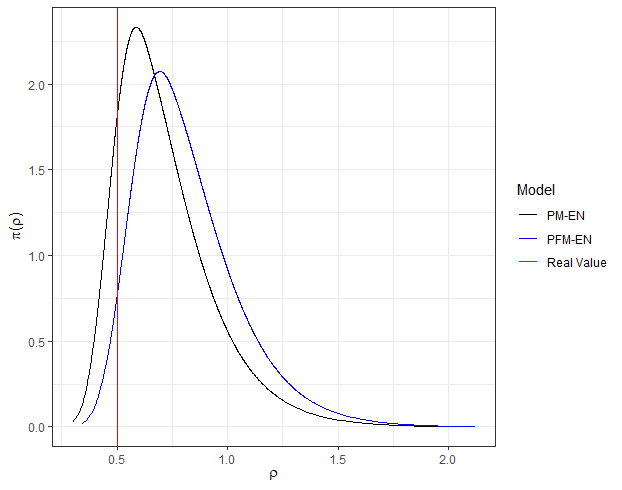}
        \caption{Spatial range posteriors for PM with EN priors.}
    \end{subfigure}\hfill%
    \begin{subfigure}{0.5\linewidth}
        \includegraphics[width=\linewidth]{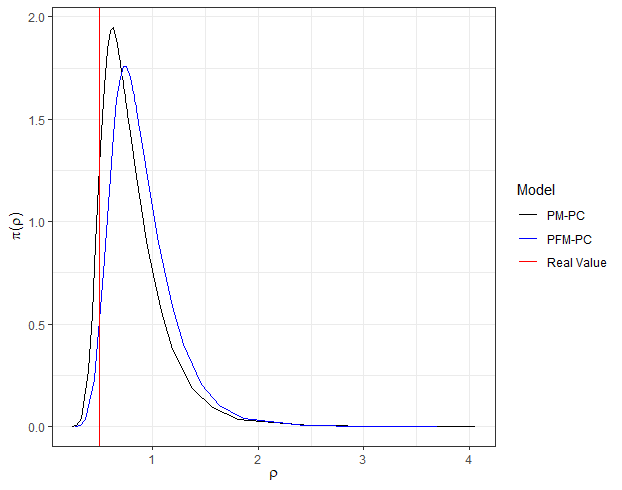}
        \caption{Spatial range posteriors for PM with PC-priors.}
    \end{subfigure}\hfill%
    
    \caption{\color{blue} Comparative example for posterior distributions of the preferential model (PM) depending on the priors used for the range and the standard deviation of the spatial effect.}
    \label{fig:PMPosteriorsExample}
\end{figure}

For hyperparameters, the distinction between models with and without feedback isn't readily apparent. However, variations are observed based on the type of prior employed and which of its moments will be updated subsequently. Notably, when we use the Normal distribution for the log  transformation of the spatial hyperparameters yields more precise values, whereas the PC-prior demonstrates greater variance, resulting in a less accurate alignment with the simulated value. Figure \ref{fig:PMPosteriorsExample} illustrates the feedback's effect on two elements of the PM. A more comprehensive comparison of posterior distributions for this instance can be found in the Appendix.

\subsection{\color{blue} Analysis and comparison of posterior distributions}

\begin{figure}[h!]
    \centering
    \includegraphics[width=\linewidth]{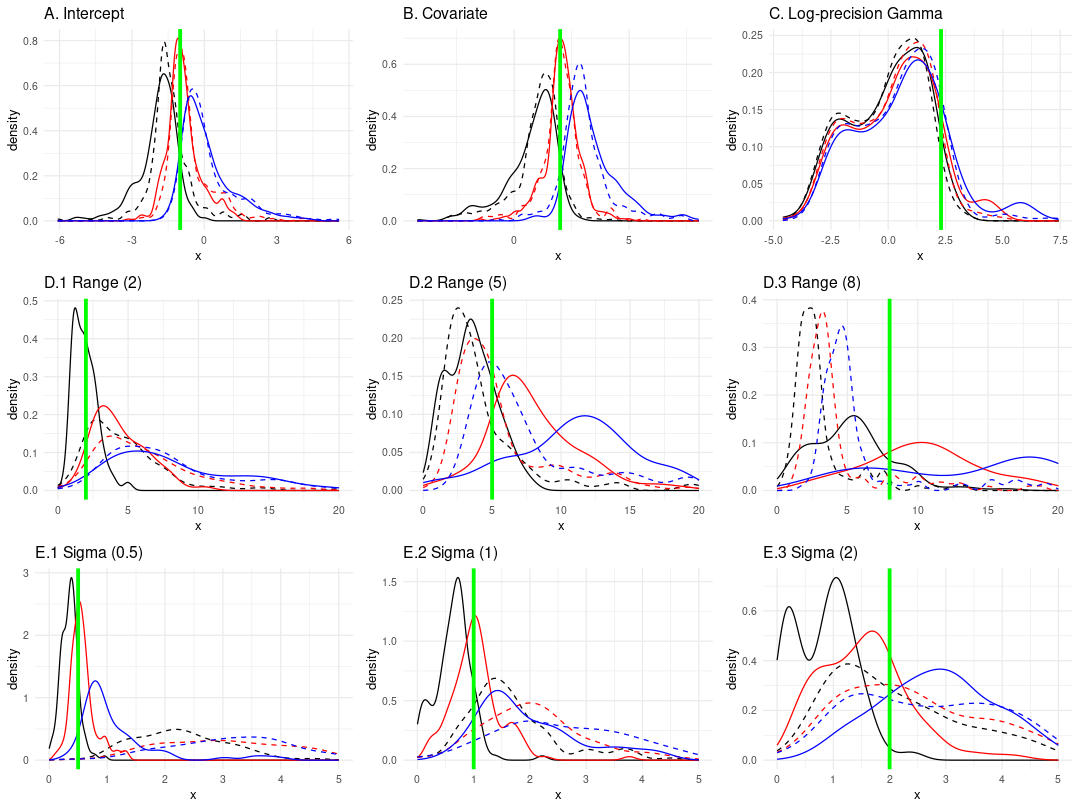}
    \caption{\color{blue} Distribution of the values of the three quantiles for the base geostatistical model (solid line) and the feedback geostatistical model (dashed lines) when we have balanced or symmetric sample sizes. The distribution of the values for the quantile $q_1$ is drawn in black, for the second quantile $q_2$ the lines are in red, for the third quantile $q_3$ is in blue, and in green the real values, we indicate the real values for the parameters and hyperparameters.}
    \label{fig:posterior_qr_symm}
\end{figure}

\begin{figure}[h!]
    \centering
    \includegraphics[width=\linewidth]{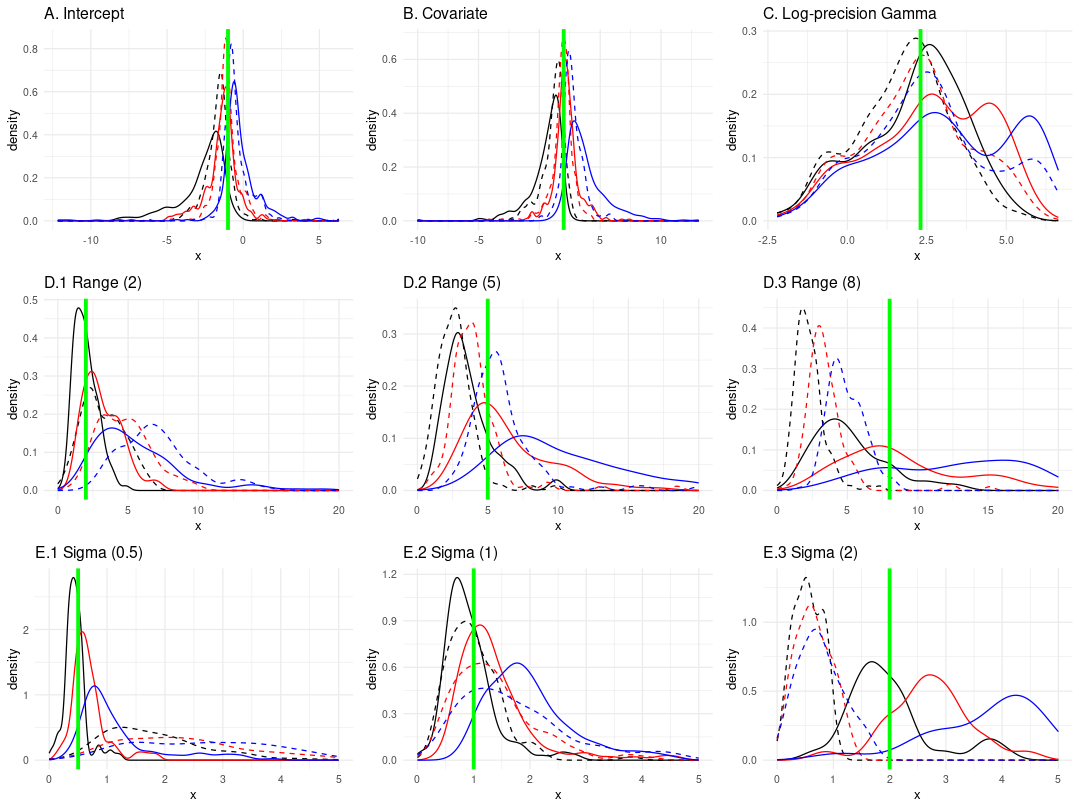}
    \caption{\color{blue} Distribution of the values of the three quantiles for the base preferential model (solid line) and the feedback preferential model (dashed lines) when we have balanced or symmetric sample sizes. The distribution of the values for the quantile $q_1$ is drawn in black, for the second quantile $q_2$ the lines are in red, for the third quantile $q_3$ is in blue, and in green the real values, we indicate the real values for the parameters and hyperparameters.}
    \label{fig:posterior_qp_symm}
\end{figure}

\begin{figure}[h!]
    \centering
    \includegraphics[width=\linewidth]{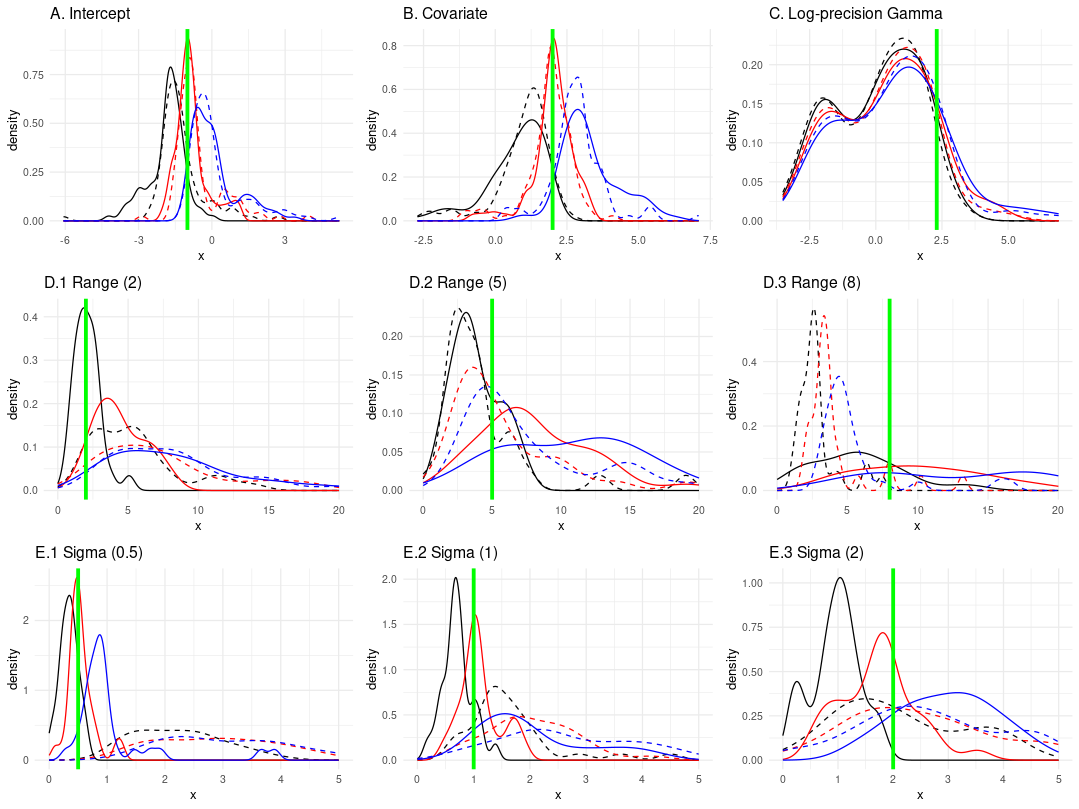}
    \caption{\color{blue} Distribution of the values of the three quantiles for the base geostatistical model (solid line) and the feedback geostatistical model (dashed lines) when we have asymmetric sample sizes. The distribution of the values for the quantile $q_1$ is drawn in black, for the second quantile $q_2$ the lines are in red, for the third quantile $q_3$ is in blue, and in green the real values, we indicate the real values for the parameters and hyperparameters.}
    \label{fig:posterior_qr_asymm}
\end{figure}

\begin{figure}[h!]
    \centering
    \includegraphics[width=\linewidth]{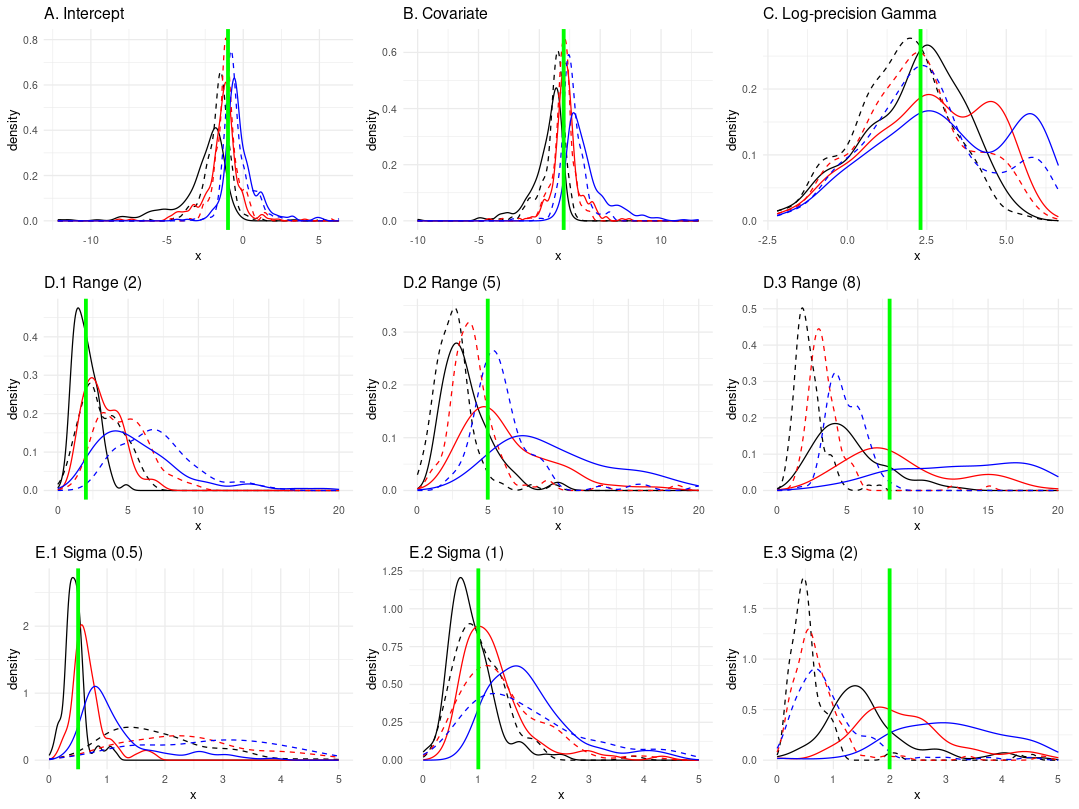}
    \caption{\color{blue} Distribution of the values of the three quantiles for the base preferential model (solid line) and the feedback preferential model (dashed lines) when we have asymmetric sample sizes. The distribution of the values for the quantile $q_1$ is drawn in black, for the second quantile $q_2$ the lines are in red, for the third quantile $q_3$ is in blue, and in green the real values, we indicate the real values for the parameters and hyperparameters.}
    \label{fig:posterior_qp_asymm}
\end{figure}

{\color{blue} In our analysis of the posterior distributions, we assess the following three quantiles: $q_1=0.025$, $q_2=0.5$ (the mean), and $q_3=0.975$. While $q_2$ gives insight into the central tendency of the distribution, $q_1$ and $q_3$ offer an understanding of the variability at the extremes, in conjunction with the replicates of the posterior distributions, which are associated with the fixed parameters and hyperparameters of the model.

In Figures \ref{fig:posterior_qr_symm} and \ref{fig:posterior_qp_symm}, we illustrate the distribution of the values for the three quantiles, comparing the values of the quantiles of the base and feedback geostatistical model and the base and feedback preferential models for the symmetric or balanced data sizes. Meanwhile, in the Figures \ref{fig:posterior_qr_asymm} and \ref{fig:posterior_qp_asymm} we show the same results as for the previous figures but related to the asymmetric data sizes.

The distributions of these quantiles for the fixed parameters, when considering the entire set of symmetric simulations, show a consistent pattern in relation to the $q_2$ quantile, with no noticeable difference between feedback and non-feedback models. However, the $q_1$ and $q_3$ quantiles tend to closely match the mean values for the feedback model. This indicates that the feedback protocol enhances accuracy by decreasing uncertainty in the estimates. This improvement is particularly clear in the case of unbalanced samples, when a low-data model is fed back to the posteriors of a big-data model.

Regarding the hyperparameters, the results in the distribution of the quantiles does not seem to consistently outperform or improve the estimation provided by the feedback models, whether in the balanced or unbalanced cases.
}

\subsection{\color{blue} Out-of-sample analysis}

{\color{blue} The assessing of the out-of-sample analysis is done by means of the Root Mean Squared Error (RMSE) and the bias. The RMSE is defined as
\begin{equation}
    RMSE = \sqrt{\frac{1}{n}\sum^n_{i=1} (y_i - \hat{y}_i)^2},
\end{equation}
while the bias is defined for the global predictive results as
\begin{equation}
    bias = \frac{1}{n}\sum_{i=1}^n |y_i-\hat{y}_i|\,,
\end{equation}
being (in both formulae) $n$ the number of simulated values, $y_i$ the simulated values, and $\hat{y}_i$ the predicted values generated by the model based on a sample set different from the one used for calculating the RMSE. 

In Figure \ref{fig:rmse_global}, we illustrate the proportion of models that have lower global RMSE compared with the alternative set of models. From the figure, we can see that $72\%$ of preferential models under feedback outperform the geostatistical base models, and $71\%$ outperform the base preferential models in terms of RMSE. The global results show that the preferential model under feedback tends to perform better than the alternative models. However, based on the results, it would be advisable to perform the feedback process for the geostatistical model as well, as the geostatistical feedback model outperforms the RMSE of the geostatistical base model by $68\%$. Moreover, in Figure \ref{fig:rmse_densities} and Table \ref{tab:summary_rmse_densities}, we present the density of the distribution of proportions relative to the global estimates shown in Figure \ref{fig:rmse_global}, comparing the following pairs of models: (A) geostatistical feedback/geostatistical base, (B) preferential base/geostatistical base, (C) preferential feedback/geostatistical base, (D) preferential base/geostatistical feedback, (E) preferential feedback/geostatistical feeback and (F) preferential feedback/preferential base. Specifically, Table \ref{tab:summary_rmse_densities} provides key statistics for assessing centrality and deviation. Additionally, it includes the values of five quantiles to quantitatively illustrate the variability of these proportions and, consequently, the distribution in which a model exhibits greater or lesser RMSE compared to the one being evaluated. 

In Figure \ref{fig:bias_global}, we illustrate the proportion of models that have lower global bias compared with the alternative set of models. The results demonstrate that the feedback process generally leads to lower biases in the models. This effect is especially notable for the geostatistical feedback model. Specifically, we observe that the bias in the predicted values for the geostatistical feedback model is lower in $90\%$ of cases compared to the geostatistical base model, $95\%$ of cases compared to the preferential base model, and $80\%$ of cases compared to the preferential feedback model.

\begin{figure}[h!]
    \centering
    \includegraphics[width=0.95\linewidth]{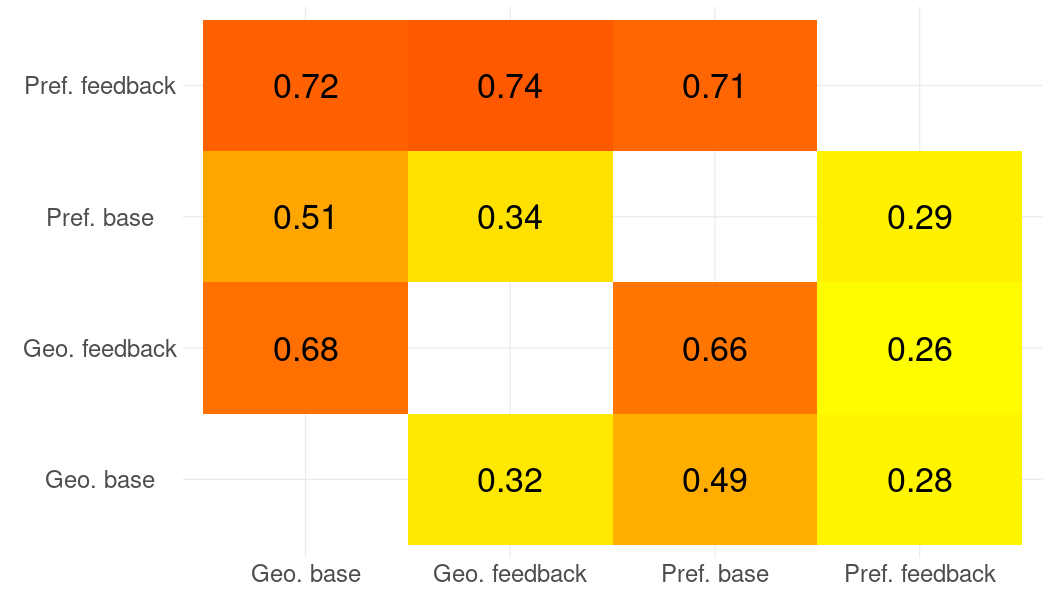}
    \caption{\color{blue} The proportion of models with a lower RMSE (in rows) compared with those for the same scenarios (in columns) for the balanced sample size analysis.}
    \label{fig:rmse_global}
\end{figure}

\begin{figure}[h!]
    \centering
    \includegraphics[width=0.95\linewidth]{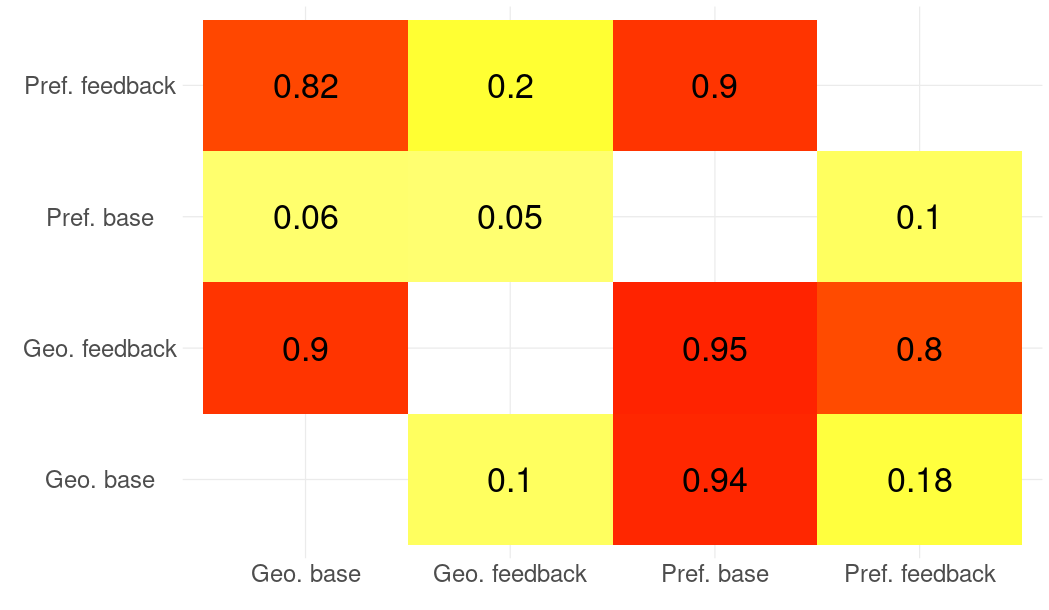}
    \caption{\color{blue} The proportion of models with a lower bias (in rows) compared with those for the same scenarios (in columns) for the balanced sample size analysis.}
    \label{fig:bias_global}
\end{figure}

\begin{figure}[h!]
    \centering
    \includegraphics[width=0.95\linewidth]{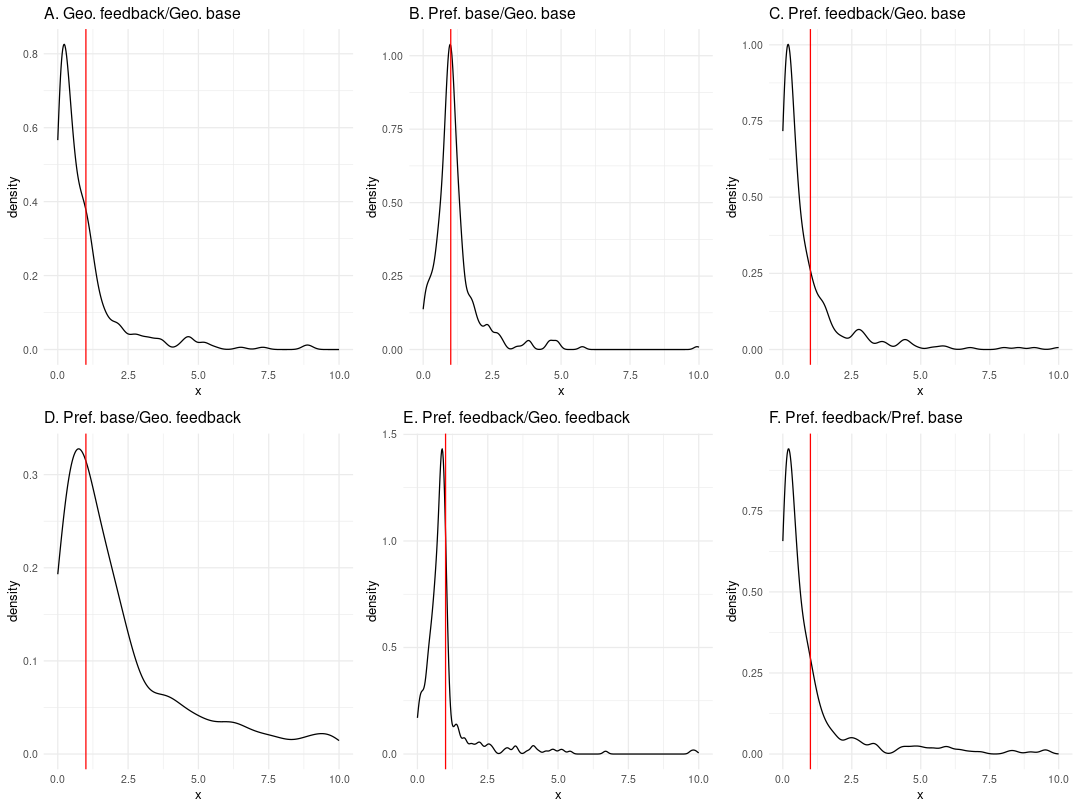}
    \caption{\color{blue} The distribution of the RMSE proportions for the different pairs of models considered throughout the study.}
    \label{fig:rmse_densities}
\end{figure}

\begin{table}
\centering
\begin{tabular}[t]{|l|r|r|r|r|r|r|r|}
\hline
  & Mean & Q. 0.025 & Q. 0.25 & Q. 0.5 & Q. 0.75 & Q. 0.975\\ \hline
A & 16.74 & 0.03 & 0.21 & 0.56 & 1.23 & 14.56\\ \hline
B & 1.23 & 0.08 & 0.74 & 0.99 & 1.31 & 4.54\\ \hline
C & 1.62 & 0.01 & 0.17 & 0.42 & 1.14 & 12.92\\ \hline
D & 20.09 & 0.07 & 0.71 & 1.63 & 4.22 & 42.01\\ \hline
E & 1.55 & 0.07 & 0.59 & 0.84 & 1.01 & 9.74\\ \hline
F & 1.79  & 0.01 & 0.18 & 0.46 & 1.13 & 13.21\\ \hline
\end{tabular}
\caption{\color{blue} Evaluation of the variability in the proportion of the global RMSE between different models. Each row contains the name of the index associated with the pair of models compared.}
\label{tab:summary_rmse_densities}
\end{table}

}

\section{\color{blue}Real data example}

{\color{blue} In this section, we present an example with real data from the fishery sciences field, in order to show the variation in the predictive results under the implementation of the different feedback procedure presented throughout the paper. In particular, two sources of information on hake abundance are available. The first source comes from a random sampling from the EVHOE scientific survey performed from 2003 to 2021. The second source of information comes from commercial data collected through observers on board in the same time interval. The region where all the locations of the sampling were performed is the southern French coast of the Atlantic Ocean, as shown in Figure \ref{fig:real_data}. 

\begin{figure}
    \centering
    \includegraphics[width=0.95\linewidth]{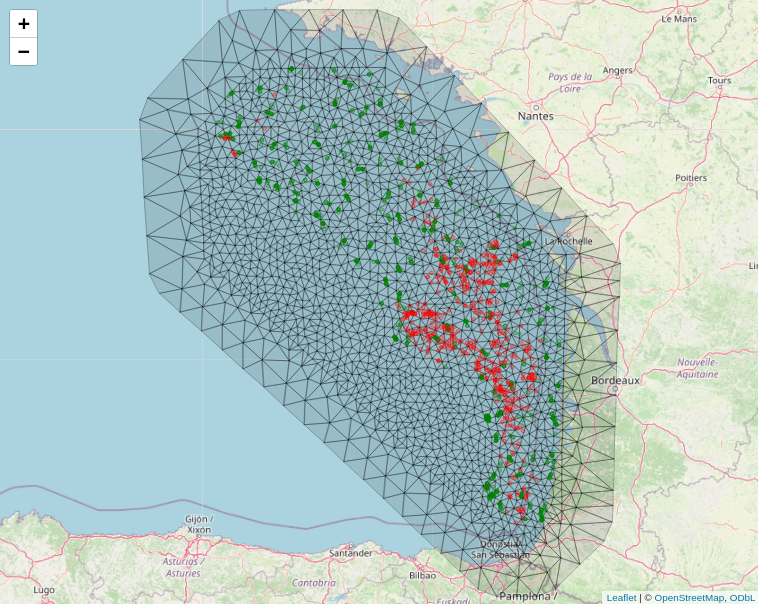}
    \caption{\color{blue} The western French coast area along with sampling locations from the scientific survey (green) and samples from the commercial surveys (red).}
    \label{fig:real_data}
\end{figure}

The scientific survey has been analysed via a Hurdle model \citep{Hurdle_Martin_2005, SpatioTemporalModelStructure_Paradinas} in order to deal with those locations with zero catches. In other words, two likelihoods are used simultaneously to model the data, a Bernouilli for the data with presence/absence of abundance and a Gamma for the data with positive non-zero abundance: 
\begin{equation}
\begin{array}{c}
     Z_i \sim \text{Bernoulli}(\pi_i),  \\
     \text{logit}(\pi_i) = \beta'_0 + \beta'_1\cdot \text{depth} + \gamma_t \cdot f_t(t_i) + \gamma_u \cdot u_i, \\
     Y_i \mid Z_i=1 \sim \text{Gamma}(\mu_i, \phi), \\
     \log(\mu_i) = \beta_0 + \beta_1\cdot \text{deph} + f_t(t_i) + u_i,
\end{array}
\end{equation}
where $\beta_0$ and $\beta_1$ are coefficients associated with the intercept and the explanatory variable depth. The term $f_t(t_i)$ is associated with each year $t_i$, such that its prior is that of a first-order random walk (RW1) and is shared between the predictor of the Gamma distribution and the Bernoulli distribution scaled by $\gamma_t$. Finally, $u_i$ is the spatial effect term that is also shared between both linear predictors and scaled by $\gamma_u$.

With respect to the commercial survey, we have used a preferential model  \citep{AccountingPreferentialSampling_Pennino} incorporating identical components for the two linear predictors, with the distinction that here we use a likelihood for the Log-Gaussian Cox process and a Gamma likelihood for the abundance data. The model is expressed as follows:
\begin{equation}
\begin{array}{c}
     Y_i \mid s_i \sim \text{Gamma}(\mu_i, \phi), \\
     \log(\mu_i) = \beta_0 + \beta_1\cdot \text{deph} + f_t(t_i) + u_i, \\
     s_i \sim \text{LGCP}(\lambda_i),  \\
     \log(\lambda_i) = \beta'_0 + \beta'_1\cdot \text{deph} + \gamma_t \cdot f_t(t_i) + \gamma_u \cdot u_i, \\
\end{array}
\end{equation}
where we have the same components with the same meaning as for the model related to the scientific survey.

After fitting both models, Figure \ref{fig:predictive_results} shows the predictive results provided by the base models and the feedback models. In this figure, we can observe that the predictions between the base and feedback models are similar. Nonetheless, discernible differences emerge in the spatial patterns, notably within the preferential model.

\begin{figure}[h!]
    \centering
    \includegraphics[width=0.95\linewidth]{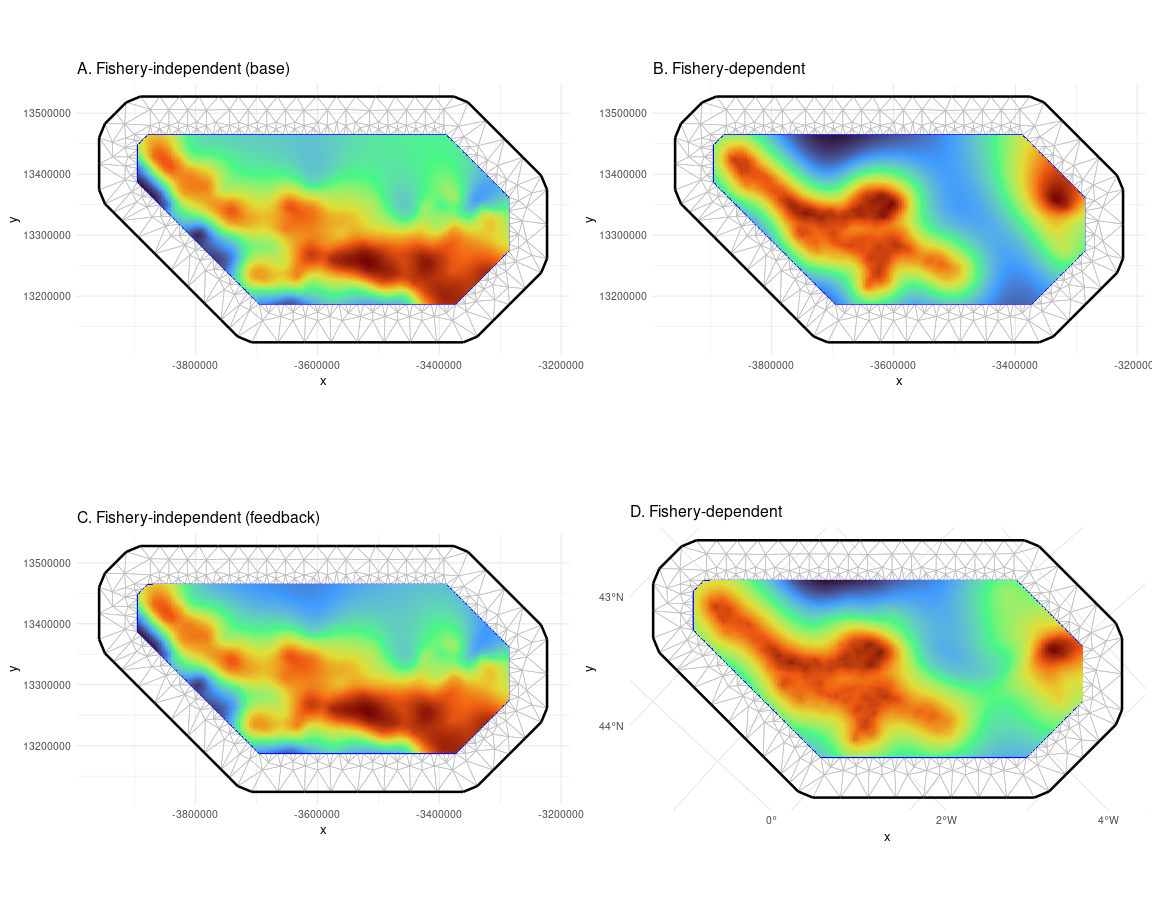}
    \caption{\color{blue} Predictive maps (temporally aggregated) obtained by the Hurdle model without feedback and with feedback (top-left and bottom-left respectively), and by the preferential model without feedback and with feedback (top-right and bottom-right respectively).}
    \label{fig:predictive_results}
\end{figure}

}

\section{\color{blue} Conclusions}

SDMs are a widely used tool for analyzing {\color{blue} spatially} distributed data and, as we have pointed out, there are scenarios in which several sources of information are available for the same phenomenon. Thus, we have analysed a particular case in which we achieve to share information between {\color{blue} geostatistical} and preferential models in both directions. However, we have also provided a generalization of the Bayesian feedback protocol applied, allowing to feed back any kind of models that fall inside of the designed framework. {\color{blue} It is important to note that if the update of all information is performed (that is, fixed parameters, random effects, and hyperparameters), then the result should not depend on the direction in which the procedure is carried out. However, in our case, we have not updated the random effects and have adhered exclusively to the fixed effects and hyperparameters.}

{\color{blue} In this paper we have used two ways to perform the feedback between the posterior and prior distributions of two given models: full update and updating by moments. Their use depends on the possibility to perform one or the other. But we have also compared two different ways of providing feed back between the two models, one geostatistical and the other preferential. In the resulting four situations have being evaluated by means of the residual analysis and commented in terms of the proportion of RMSE and bias.}

The main conclusions can be summarised as follows. The feedback process generally improves the accuracy of the latent field parameters, while this is not always true for the hyperparameter field. {\color{blue} The feedback procedure appears to enhance the prediction of preferential and geostatistical models, with the preferential feedback model showing the best performance in terms of RMSE}. Additionally, feedback increases the robustness of the preferential model, which may encounter computational issues in certain scenarios \citep{conn2017confronting}. This study resolved this robustness issue in preferential modelling by adjusting some internal parameters of \textbf{INLA}, while feedback-ed models required no adjustments.

In light of the above findings, it can be concluded that providing feedback to the preferential model generally leads to improvements in result accuracy, computational efficiency, and robustness, the latter being a general concern in preferential models \citep{conn2017confronting, GeostatisticalPreferentialSampling_Diggle}. While this improvement tends to be more pronounced compared to its alternative, the difference in accuracy is negligible when more precise results are not obtained. In contrast, the other two improvements consistently occur, thereby reducing modelling time and streamlining computational resolution. Therefore, it is recommended to use a simpler model to provide feedback into a more complex model, enhancing computational stability, reducing computational costs, and potentially yielding more accurate outcomes.

{\color{blue} To the best knowledge of the authors, this is the first study that use elicitation over a preferential model. The current increase in the number of opportunistic data sources available in ecology makes this study particularly interesting as it provides a relatively simple approach to enhance their use in terms of robustness and accuracy. Other studies have demonstrated that elicitation helps reduce uncertainty in species habitat predictions implementing Bayesian models and some of them with INLA, but generally combining expert information to improve geostatistical models \citep{pearce2001incorporating, di2018expert}. Despite the differences in the modeling approach, our findings align well with these earlier studies. The combination of multiple data sources helps to increase the robustness and reliability of the results. This is very important in contexts such as fisheries, because in order to achieve efficient management of marine resources there is often a need for alternative methods to increase confidence in fisheries-dependent data \citep{vanhatalo2014catch}. In summary, our results are encouraging for wider use of these methods and could be adapted to larger scale applications; such as environmental management. In fact, today most climate change projections are made on a global scale and it is difficult to translate them to a regional scale. Elicitation could help these priority objectives of the Ocean Decade.}

\section{\color{blue} Code and data availability}

{\color{blue} The link to the code and the data used can be found in the following GitHub repository: \url{https://github.com/MarioFigueiraP/Feedback_code_article}.}

\cleardoublepage

\section*{Acknowledgments}

This study is part of the ThinkInAzul program and is financed by the MCIN with funds from the European Union (NextGenerationEU-PRTR-C1711) and by the Generalitat Valenciana GVA-THINKINAZUL/2021/021. DC acknowledges Grant CIAICO/2022/165 funded by Generalitat Valenciana. 
XB, DC and ALQ thank support by the grant PID2022-136455NB-I00, funded by Ministerio de Ciencia, Innovación y Universidades of Spain (MCIN/AEI/10.13039/501100011033/FEDER, UE) and the European Regional Development Fund. MGP also thanks the project FRESCO (PID2022-140290OB-I00), funded by Ministerio de Ciencia, Innovación y Universidades of Spain.  

\section*{Conflict of Interest}

The authors declare no conflict of interest.

\cleardoublepage

\bibliographystyle{elsarticle-harv} 
\bibliography{article_feedback.bib}

\end{document}